\documentclass[twocolumn,showpacs,amsfonts,amssymb,amsmath,floats,superscriptaddress,aps]{revtex4-1}
\usepackage{graphicx}
\usepackage{dcolumn}
\usepackage{multirow}

\usepackage{braket}

\def\w{\omega}

\usepackage{lineno}
\usepackage{upgreek}
\usepackage{bbold}
\usepackage[normalem]{ulem}

\usepackage{xcolor}
\usepackage{placeins}

\def\doubleunderline#1{\underline{\underline{#1}}}

\renewcommand{\vec}[1]{\mathbf{#1}}
\usepackage[colorlinks]{hyperref}
\usepackage{mathtools}

\DeclareGraphicsExtensions{.pdf,.eps}
\graphicspath{{figs/}}

\begin{document}
\title{Nuclear spin noise in the central spin model}

\author{Nina Fr\"ohling}
\address{Lehrstuhl f\"ur Theoretische Physik II, Technische Universit\"at Dortmund,
Otto-Hahn-Stra{\ss}e 4, 44227 Dortmund, Germany}

\author{Frithjof B. Anders}
\address{Lehrstuhl f\"ur Theoretische Physik II, Technische Universit\"at Dortmund,
Otto-Hahn-Stra{\ss}e 4, 44227 Dortmund, Germany}

\author{Mikhail Glazov}
\address{Ioffe Institute, Polytechnicheskaya 26, 194021 St. Petersburg, Russia}

\date{\today}

\begin{abstract}

We study theoretically the fluctuations of the
nuclear spins in quantum dots employing
the central spin model which accounts for the hyperfine interaction of the
nuclei with the electron spin. These
fluctuations are calculated
both with an analytical approach
using homogeneous hyperfine couplings (box model) and
with a numerical simulation
using a distribution of hyperfine coupling constants.
Both approaches are in good agreement. The box model serves
as a benchmark with low computational cost
that explains the basic features of the nuclear spin noise well.
We also demonstrate that  the nuclear spin noise spectra
comprise
a two-peak structure centered at the nuclear Zeeman frequency in  high magnetic fields
with the shape of the spectrum controlled by the distribution of the hyperfine constants.
This allows for a direct access to this distribution function
through nuclear spin noise spectroscopy. 
\end{abstract}

\setcounter{page}{0}
\pagenumbering{arabic}

\maketitle

\section{Introduction}

The nuclear spin fluctuations~\cite{Bloch46} play an
important role in the spin dynamics in semiconductor nanosystems~\cite{dyakonov_book}.
In particular, these fluctuations provide an efficient mechanism of the electron spin decoherence in quantum dots~\cite{PhysRevB.65.205309,PhysRevLett.88.186802}
due to the hyperfine interaction with electron spin.
The nuclear spin-spin correlation function also
controls the ``warm-up'' of the nuclear spin system in the alternating magnetic field~\cite{opt_or_book}.

The spin noise technique \cite{aleksandrov81} provides
a method for almost non-perturbative studies of the spin dynamics in thermal equilibrium or close to equilibrium conditions. 
In this technique a linearly polarized light beam propagating through the transparency region of the sample experiences a random fluctuation
of its polarization plane orientation caused by stochastic fluctuations of the magnetization of the media via the Faraday effect. 
First introduced 
to atomic systems~\cite{aleksandrov81,PhysRevLett.80.3487,Crooker_Noise}, this technique has 
successfully been
applied to semiconductors and semiconductor nanosystems~\cite{Oestreich_noise}, see Refs.~\cite{Oestreich:rev,Zapasskii:13,2016arXiv160306858S} for reviews. 
Since the optical response of a semiconductor is dominated, as a rule, by the electronic excitations, 
the spin noise spectroscopy technique has actively been used to study the electron spin fluctuations. 
The nuclear spin fluctuations 
determine the electron spin noise spectra in quantum dots~\cite{Glazov2012,Glasenapp2016}. 
The nuclear spin dynamics was revealed
in the specially designed~\cite{PhysRevB.91.205301} experiments:
The nuclear spins were first polarized, 
and afterwards the nuclear spin relaxation has been monitored via the electron spin noise~\cite{ryzhov15,Ryzhov:2016aa}.

However, the nuclear spins also provide a sizable 
contribution to the Faraday effect in semiconductors~\cite{artemova85,PhysRevLett.111.087603}. 
Recently, their fluctuations have been observed by the spin noise technique in the GaAs sample with donor-bound electrons~\cite{2015arXiv150605370B}. 
This motivates us to study theoretically the nuclear spin fluctuations for a system where a charge carrier, e.g., 
an electron, (i) is localized in a quantum dot or (ii) is bound to a neutral donor 
and interacts 
with a large number of nuclear spins in its vicinity. The coupled spin dynamics is accounted for by 
the central spin model that
has been studied previously  with the emphasis on the central spin 
(i.e., charge carrier)
dynamics both for systems in 
equilibrium \cite{PhysRevB.89.045317, PhysRevLett.115.207401, PhysRevB.76.014304, PhysRevB.96.045441} and for periodically pulsed systems \cite{Petrov2012, PhysRevB.96.205419, PhysRevB.85.041303}. 
While the interactions between the
nuclear spins are much smaller than the dominating hyperfine interaction and, therefore, are often neglected, 
the charge-carrier spin dephasing in a bath of interacting spins has also been investigated \cite{PhysRevB.74.195301, PhysRevB.74.035322, PhysRevLett.102.057601}. However, nuclear spin fluctuations have not been calculated within the central spin model to the best of our knowledge.  

In this paper, we present results for the nuclear spin correlation function and the nuclear spin noise 
within the central spin model including the Zeeman effect caused by the external magnetic field. We apply two complementary approaches based on (i) an analytical solution within the ``box'' model approach where all hyperfine coupling constants are assumed to be equal and (ii) a direct numerical procedure 
for a distribution 
of the hyperfine coupling constants. 

In the box model we can treat realistically large nuclear bath sizes at a 
relatively small computational expense. Furthermore,
simple and physically transparent results are obtained in the important limits of low and high magnetic fields, which can be readily compared with experimental data. The box model calculations also serve as a test-bed for the advanced numerical calculations which are essentially restricted by small sizes of the nuclear spin bath, but allow us to account for the 
distribution of the hyperfine coupling constants. Demonstrating the validity and good agreement between the approaches, we also demonstrate how the distribution of the hyperfine coupling constants can be determined from the nuclear spin noise spectrum.  

The paper is organized as follows: In Sec.~\ref{sec:mode} we briefly formulate the central spin model and introduce the notations. In Sec.~\ref{sec:SNS_box} we present the analytical solution of the central spin model within the ``box'' model. 
The results for the spin noise spectra are covered in Sec.~\ref{sec:sns}. Section~\ref{sec:full} contains 
the numerical data for the central spin model with a distribution of 
the hyperfine coupling constants. A brief summary of results is given in Sec.~\ref{sec:concl}.

\section{Model}\label{sec:mode}


In order to describe the
nuclear spin fluctuations in electron doped quantum dots, we have to model all relevant interactions. The system is dominated by the Fermi contact hyperfine interaction between the electron spin $\bold{S}$ and the bath of $N$ nuclear spins, $\bold{I}_k$, $k=1, \ldots, N$, with the characteristic time scale of several nanoseconds in typical epitaxial III-V quantum dots~ \cite{PhysRevLett.109.166605, PhysRevLett.115.207401}. The dipole-dipole interaction between the nuclear spins 
is five orders of magnitude weaker, only contributing to the dynamics on the time scale of $100$~$\upmu \text{s}$ \cite{PhysRevB.65.205309}. The strength of quadrupolar interaction, caused by strain fields in the material, is highly dependent on the material. It was measured to be in the order of $\sim 100$~ns \cite{PhysRevLett.115.207401} for InAs quantum dots. In this work we will focus on the dominant hyperfine interaction and exclude the much weaker quadrupolar and dipole-dipole interaction.
This simplification leads to the central spin model Hamiltonian $H_{\text{CSM}}$  ~\cite{Gaudin1976} 
\begin{align}
\begin{split}\label{CSM}
\hat H_{\text{CSM}} = & \sum_{k=1}^N  A_k \bold{I}_k\cdot\bold{S}
+ \sum_{k=1}^N g_{k}\mu_{N}\vec{B}\cdot\vec{I}_k +g \mu_{\text{B}}\vec{B}\cdot\vec{S}.
\end{split}
\end{align}
Here $A_k$ are the hyperfine coupling constants,
$\bold{B}$ is the external magnetic field, $g$ is the charge carrier $g$-factor, $(-g_{k})$ are the nuclear 
$g$-factors, and $\mu_{N}$ and $\mu_{\rm B}$ are the nuclear and Bohr magnetons, respectively. 

The hyperfine coupling results in the intrinsic time scale $T^\ast$ defined via
\begin{align}
\label{Tstar}
\frac{1}{{T^{\ast}}^2}= \sum_{k=1}^N \frac{4I_k(I_{k}+1) }{3} A_k^2.
\end{align}
In the following
we assume that all nuclei have the same spin $I_{k}\equiv I_0$, put $\hbar=1$ and use $T^\ast$ as a unit of time. Correspondingly, we redefine the dimensionless Hamiltonian $T^{\ast} \hat H_{\text{CSM}} \to \hat H_{\text{CSM}}$, the dimensionless hyperfine coupling constants $a_k = T^{\ast}A_k$ and the dimensionless
external magnetic field $\vec b = T^{\ast}g \mu_{\text{B}}\bold{B}$. 
It is also convenient to introduce the ratio between the nuclear and the electron Zeeman splitting
\begin{align}
z_k = \frac{g_{{k}}\mu_{N}}{g\mu_{\text{B}}}.
\end{align}
Finally, the dimensionless Hamiltonian takes the form
\begin{align}
H_{\text{CSM}}=\vec S \cdot\left(\vec b +\sum_{k=1}^N  a_k \bold{I}_k\right)+ \vec b\cdot \sum_{k=1}^N  z_k \bold{I}_k.
\end{align}
The parameter $T^\ast$ in Eq.~\eqref{Tstar} is on the order of several nanoseconds, and the hyperfine interaction strength
corresponds to the temperatures of the order $10\,\text{mK}$ \cite{doi:10.1063/1.1850605} in self-assembled InGaAs quantum dots.
The standard spin-noise experiments on epitaxial III-IV quantum dots are performed at temperatures of about $4\ldots 6$~K, which allows us to disregard the intricate physics of nuclear spin polaron formation and the nuclear self-polarization occurring at much lower temperatures~\cite{Merkulov:1998aa,PhysRevB.95.245209,dp_72_self,Korenev:1999aa}. Therefore, we employ the  
high-temperature limit and assume that all states are equally occupied.

The strength of the hyperfine coupling $A_k$ is proportional to the probability of the electron to be 
at the position of the $k$th nucleus, $|\psi_e(\bold{R}_k)|^2$.
The envelope function of
the electron wave function in a $d$-dimensional  quantum dot can be written as
\begin{align}
\label{wavefunction}
\psi_e(\bold{r})= CL_0^{-{d}/2}\exp \left(-\frac{|\bold{r}|^m}{2L_0^m}\right) .
\end{align}
Here $C$ is the dimensionless normalization constant, $L_0$ is the effective localization radius, and the parameter $m$ determines the shape of the wave function: $m=1$ corresponds to a hydrogen-like electron wave function, while $m=2$ describes a Gaussian wave function.
Several different distributions have been used in simulations of the central spin model \cite{PhysRevB.89.045317, PhysRevB.70.195340, PhysRevB.88.085323}. However, the exact distribution of the coupling constants does not influence the short time (on the scale of $T^*$) dynamics of the system.

\section{Exact Solution for the Box Model}
\label{sec:SNS_box}

\subsection{Energy spectrum and eigenstates}

We start with the simplest possible case,
the so-called box model, where the hyperfine coupling constants $A_k\equiv A_0$ are all the same.
Furthermore, we assume that the $g$-factors for all nuclei are identical, i.e., $z_k\equiv z$
and assume that each nuclear spin is $I_0=1/2$. 
Then it follows from Eq. \eqref{Tstar} that the characteristic time scale is $T^{\ast}= 1/A_0\sqrt{N}$,
 and the dimensionless coupling constant $a_0=1/\sqrt{N}$. 
Even though the number of nuclear spins is macroscopic, $N\rightarrow\infty$
in an experimental 
sample,  the number of nuclear spins effectively coupled to the central spin
with a significant coupling constants, $N_{\text{eff}}$, remains 
finite and can be estimated by the characteristic length scale of the localized electronic wave
function. This is because only the nuclei within the electron localization volume effectively interact with the charge carrier. In experiments $N\to N_{\text{eff}}\approx 10^5$ is a typical value for quantum dot samples.

Defining the total bath spin $\bold{I}=\sum_k \bold{I}_k$ results in the Hamiltonian
\begin{align}
H = hS_z + zhI_z + a_0 \left[I_zS_z+\frac{1}{2}(I^+S^-+I^-S^+)\right].
\label{box_model_nodim}
\end{align}
for an
external field $\mathbf b=h\mathbf{e}_z$.
Equation~\eqref{box_model_nodim} can also be  recast into the form
\begin{align}
\label{box_model_nodim:total}
H = hS_z + zh L + \frac{a_0}{2}(F^2-I^2-S^2),
\end{align}
where the total spin $\bold{F}=\bold{I}+\bold{S}$ and $L$ is the $z$ component of the total nuclear spin. 

In the special case of $h=0$, the Hamiltonian~\eqref{box_model_nodim:total} can be diagonalized via 
addition of the angular momenta
resulting in an energy spectrum of
the form~\cite{0953-8984-15-50-R01,1742-5468-2007-06-P06018,PhysRevB.76.014304,Kozlov2007}:
\begin{equation}
\label{h=0case}
E_{F,F_z, I} = \begin{cases}
      \frac{a_0}{2}I, & \text{if}\ F=I+1/2, \\
      -\frac{a_0}{2}(I+1), & \text{if}\ F=I-1/2.
    \end{cases}
\end{equation} 
In order to obtain the solution 
for a finite magnetic field, 
it is convenient to introduce the nuclear spin states $\ket{I,L}$, where $L=-I,\ldots, I$ is the $z$-component of the nuclear spin and use the basis set 
\begin{align}
\begin{split}
B = & \{\ket{I, -I}\ket{\downarrow},  \ket{I, -I}\ket{\uparrow}, \ket{I, -I+1}\ket{\downarrow},\\
  &\ket{I, -I+1}\ket{\uparrow},..., \ket{I, I}\ket{\downarrow},  \ket{I, I}\ket{\uparrow}\},
  \end{split}
  \label{BasisB}
\end{align}  
where $\ket{\uparrow}$, $\ket{\downarrow}$ denote the electron spin states with $S_z=\pm 1/2$, respectively. In the basis~\eqref{BasisB} the Hamiltonian matrix can be expressed in the block-diagonal form as~\cite{Kozlov2007}
\begin{align}
H \!\! = \!\!
\begin{pmatrix}
E_{+}(I, {-I})	& &	& \dots	 & &0      \\
	& M_I^{-I+1} 	& & \dots  & &	  \\
\vdots	&& M_I^{-I+2} 	&  & & \vdots \\
\vdots	&& 	& \ddots && \vdots \\
\vdots	&& 	&  & M_I^{I}&\vdots \\
0 	& & \dots & 	 && E_{-}(I, {I+1})
\end{pmatrix}.
\label{Hmatrix}
\end{align}
The states $\ket{I, -I}\ket{\downarrow}$  and $\ket{I, I}\ket{\uparrow}$ are decoupled since electron-nuclear flips are impossible. Correspondingly, the formally defined quantities $E_{+}(I, -I)$ and $E_-(I,I+1)$ in the matrix~\eqref{Hmatrix} are scalars of the value
\begin{align}
&E_{+}(I, {-I})= \bra{I, -I}\bra{\downarrow} H \ket{\downarrow}\ket{I, -I} = \frac{a_0 I - h}{2}-zhI,\\
&E_{-}(I, {I+1}) = \bra{I, I}\bra{\uparrow} H \ket{\uparrow}\ket{I, I} = \frac{a_0 I + h}{2}+zhI,
\label{EIpm}
\end{align}
and $M_I^L$ are $2\times2$ matrices in the basis $\{\ket{I,L}\ket{\downarrow}, \ket{I,L-1}\ket{\uparrow}\}$,
\begin{align}
M_I^L = 
\begin{pmatrix}
-\frac{h+a_0L}{2}+zhL & \zeta_I^L\\
\zeta_I^L & \frac{h+a_0(L-1)}{2}+zh(L-1),
\end{pmatrix}
\end{align}
with
\begin{align}
\zeta_I^L = \frac{a_0}{2}\sqrt{I(I+1)-L(L-1)}.
\end{align}
Diagonalization of matrix~\eqref{Hmatrix} results in the eigenvalues 
\begin{align}
E_{\pm}(I, L) = -\frac{a_0}{4}\pm \frac{1}{2}\sqrt{\left(\frac{a_0}{2}(2L-1) + h(1-z)\right)^2+4\left(\zeta_I^{L}\right)^2}.\label{eigenvalues:total}
\end{align}
and the eigenstates
\begin{align}
\label{eigenvalues:total1}
\begin{split}
|I, &{L}, -\rangle = \\& \mathcal A_{I, {L}, -}\ket{I, {L}}\ket{\downarrow}+\mathcal B_{I, {L}, -}\ket{I, {L-1}}\ket{\uparrow},\\
|I, &{L}, +\rangle = \\& \mathcal A_{I, {L}, +}\ket{I, {L}}\ket{\downarrow}+\mathcal B_{I, {L}, +}\ket{I, {L-1}}\ket{\uparrow}.
\end{split}
\end{align}
Here, the pre-factors $\mathcal A_{I, {L}, \pm}$ and $\mathcal B_{I, {L}, \pm}$, 
\begin{align}\label{AB}
\begin{split}
\mathcal A_{I, {L}, \pm} = \frac{2\zeta_I^{{L}}}{\sqrt{(h+a_0L+2E_{{L}, \pm}-2zhL)^2+(2\zeta_I^{{L}})^2}}   \, ,\\
\mathcal B_{I, {L}, \pm} = \frac{h+a_0L+2E_{{L}, \pm}-2zhL}{\sqrt{(h+a_0L+2E_{{L}, \pm}-2zhL)^2+(2\zeta_I^{{L}})^2}},
\end{split}
\end{align}
describe the mixing of the Ising states~\eqref{BasisB} constituting the eigenstates.

The first and the last eigenstate are the Ising states with extremal values of the total spin 
component $F_z$
\begin{align}\label{eigenvalues:total2}
\begin{split}
\ket{I, {-I}, +} &= \ket{I, -I}\ket{\downarrow}\\
\ket{I, {I+1}, -} &= \ket{I, I}\ket{\uparrow}
\end{split}
\end{align}
and the corresponding eigenvalues presented in Eq. \eqref{EIpm}. 
It is convenient to extend the
Eqs.~\eqref{AB} to take into account
${L=-I, I+1}$ and define ${\mathcal{A}}_{I, {-I}, -}={\mathcal{B}}_{I, {-I}, \pm}={\mathcal{A}}_{I, {I+1}, \pm}={\mathcal{B}}_{I, {I+1}, +}=0$ as well as ${\mathcal{A}}_{I, {-I}, +}={\mathcal{B}}_{I, {I+1}, -}=1$.

Equations~\eqref{eigenvalues:total}, \eqref{eigenvalues:total1} and \eqref{eigenvalues:total2} describe the 
energy spectrum and the eigenfunctions 
of the box model accounting for Zeeman splitting of both the electron and the nuclei.
 Eq.~\eqref{eigenvalues:total} reduces to Eq.~\eqref{h=0case}
for $h=0$.
In the case of a
negligible hyperfine coupling, $a_0\to 0$, the system separates 
into decoupled non-interacting nuclei and a single electron
with the energy spectrum given by the corresponding Zeeman splittings.

\subsection{Calculation of the spin correlation functions}

Having diagonalized the system, it is possible to evaluate the dynamics of any desired observable that can be expressed in the basis $B$, Eq. \eqref{BasisB}. Here, we are specifically interested in the
autocorrelation functions of the form $C(t)=\braket{O(t)O(0)}$ with $O(t)$ 
being the operator of the electron or nuclear spin taken in the Heisenberg representation. 
Expressing the time-evolution operator of the system via the eigenvalues and eigenstates~\eqref{eigenvalues:total}, \eqref{eigenvalues:total1} and \eqref{eigenvalues:total2} we arrive at
\begin{multline}
C(t) = \frac{1}{Z} \sum_{I=0}^{N/2}N_w(I)\!\! \sum_{\substack{\sigma = \pm \\ \sigma' = \pm}}{\sum_{\substack{L=-I\\L'=-I}}^{I+1}} |\braket{I, {L}, \sigma| O |I,  {L'}, \sigma'}|^2\\ 
\times \exp\{\text{i}[E_{\sigma}(I, {L})-E_{\sigma'}(I, {L'})]t\}.
\end{multline}
Here $Z$ is the partition function, $N_w(I)$ denotes
the number of configurations of $N$ bath spins  that have the total bath spin length of $I$, and it is given by
\begin{align}
N_w(I) = \frac{(N/2-I)!(2I+1)}{N/2+I+1}.
\end{align}
We made use of the high temperature limit where $Z=2^{N+1}$ for $1/2$ bath spins. 
In the next section we present analytical and numerical results for the autocorrelation functions and spin-noise power spectra defined as
\begin{equation}
\label{Fourier:def}
(O^2)_\omega = \int_{-\infty}^{\infty} d t e^{\mathrm i \omega t} C(t),
\end{equation}
for the electron and nuclear spin components. We consider both fluctuations of the longitudinal and transversal spin components with respect to the external magnetic field orientation.

\section{Spin noise spectra}\label{sec:sns}

\subsection{Box model}

For the autocorrelation function of the electronic spin transverse to the magnetic field, one arrives at
\begin{align}
\begin{split}
\braket{S_x(t)S_x(0)}=&
\frac{1}{4Z}  \sum^{N/2}_{I=0} N_w(I)\!\!\!\!\sum_{\sigma, \sigma'=\pm} \sum_{{L=-I}}^{{I}} \mathcal A_{{I, L}, \sigma}^2 \mathcal B_{{I, L+1}, \sigma'}^2\\
&\times \cos\{[E_{\sigma}(I, {L})-E_{\sigma'}(I, {L'})]t\},
\end{split}
\end{align}
and for the autocorrelation function of the electron spin longitudinal to the magnetic field
we obtain
\begin{align}
\begin{split}
\label{eqn:23}
\langle S_z&(t)S_z(0)\rangle=
\frac{1}{4Z}  \sum^{N/2}_{I=0} N_w(I)\!\!\!\!\sum_{\sigma, \sigma'=\pm} \sum_{{L=-I}}^{{I+1}} |\mathcal B_{{I, L}, \sigma}\mathcal B_{I, {L}, \sigma'}\\&-\mathcal A_{I, {L}, \sigma}\mathcal A_{I, {L}, \sigma'}|^2
 \exp\{\text{i}[E_{\sigma}(I, {L})-E_{\sigma'}(I, {L'})]t\}.
\end{split}
\end{align}
The autocorrelation function of the total bath spin can be computed analogously, both
for the transverse spin component to the magnetic field
\begin{align}
\begin{split}
&\langle I_x(t)I_x(0)\rangle=\frac{1}{4Z}\sum^{N/2}_{I=0}N_w(I)\!\!\!\!\sum_{\sigma, \sigma'=\pm}\! 
\times \\
& \sum_{{L=-I}}^{{I}}[\mathcal A_{{I, L}, \sigma} \mathcal A_{{I, L+1}, \sigma'} \mathcal B_{{I, L}, \sigma}\mathcal B_{{I, L+1}, \sigma'}\\
&\times \sqrt{I(I+1)-({L}-1){L}}\\
&\times \sqrt{I(I+1)-({L}+1){L}}]^2\\
&\times \cos[(E_{\sigma'}({I, L+1})-E_\sigma({I, L}))t]
\label{IxIx_box}
\end{split}
\end{align}
and the longitudinal spin component to the magnetic field
\begin{align}
\begin{split}
&\langle I_z(t)I_z(0)\rangle=\frac{1}{4Z}\sum^{N/2}_{I=0}N_w(I)\!\!\!\!\sum_{\sigma, \sigma'=\pm} \sum_{{L=-I}}^{{I+1}} |{L}\mathcal A_{{I, L}, \sigma}\mathcal A_{{I, L}, \sigma'}\\&-({L}+1)\mathcal B_{{I, L}, \sigma}\mathcal B_{{I, L}, \sigma'}|^2
\exp{\{\text{i}[E_{\sigma}(I, {L})-E_{\sigma'}(I, {L'})]t\}}.
\label{IzIz_box}
\end{split}
\end{align}

We note that for the macroscopic number of nuclei the electron spin autocorrelation functions reduce to the expressions derived within the semi-classical model where the nuclear spins are considered as frozen~\cite{PhysRevB.65.205309,Glazov2012}: 
For instance,  
spin fluctuations are isotropic at $h=0$, and
Eq.\ \eqref{eqn:23} reduces to
\begin{align}
\begin{split}
\langle S_z&(t)S_z(0) \rangle = \frac{1}{2Z}\sum_{I=0}^{N/2}\frac{N_w(I)}{(2I+1)^2}\sum_{L=-I}^{I}\{(2L-1)^2\\
&+4(I+L)(I-L+1)\cos[a_0(2I+1)t/2] \}.
\end{split}
\end{align}
The direct calculation of the sum~\cite{Kozlov2007} at $N\to \infty$ yields
\[
\langle S_z(t)S_z(0) \rangle = \frac{1}{4} \left[\frac{1}{3} + \frac{2}{3}(1-t^2/4)\exp{(-t^2)} \right]
\]
in full agreement with Refs.~\cite{PhysRevB.65.205309,Glazov2012}. 
Since the total spin $\mathbf{F} = \mathbf{S} + \mathbf{I}$ 
is conserved in the absence of an external magnetic field,
the nuclear spin $\mathbf{I}$ remains constant within the tolerance of $1/\sqrt{N}\to 0$
leading to the constant correlation function
\begin{equation}
\label{auto:h=0}
\langle I_z(t) I_z(0)\rangle = \frac{N}{4}
\end{equation}
in the limit $N\to \infty$.

\begin{figure}[tbp]
\begin{center}

\includegraphics[width=0.49\textwidth,clip]{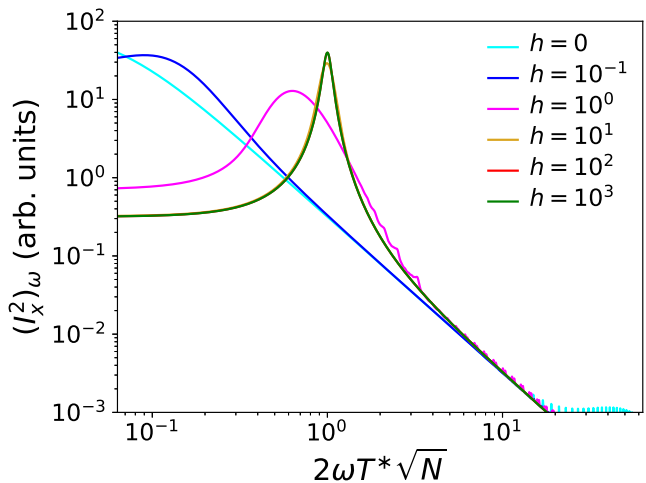}
\caption{The nuclear spin noise (SN) for different transversal external magnetic fields $h$ and a bath size of $N=1000$, computed within the box model as derived from a Fourier-transformation of 
Eq.\ \eqref{IxIx_box}. The nuclear Zeeman splitting is set to zero.}
\label{Box_Ix_noz}
\end{center}

\end{figure}

\begin{figure}[tbp]
\begin{center}
\includegraphics[width=0.49\textwidth,clip]{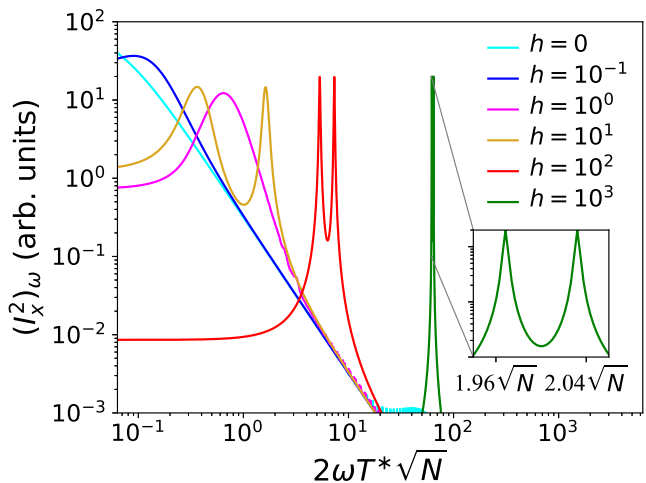}
\caption{The nuclear spin noise for different transversal external magnetic fields $h$ and a bath size of $N=1000$, computed within the box model, again derived from a Fourier transformation of Eq.\ \eqref{IxIx_box}. The ratio of nuclear to electron Zeeman splittings is $z=10^{-3}$.}
\label{Box_Ix_z}
\end{center}
\end{figure}

The nuclear spin noise 
is given by the Fourier transformation of the expression~\eqref{IxIx_box}, 
resulting
in a sum of $\delta$-function peaks centered at $$\omega_{L, \sigma, \sigma'} = \pm [E_{\sigma'}(L+1)-E_\sigma(L)].$$
Introducing additional dephasing rate $\gamma$, 
we replace $\delta$-functions by a broadened Lorentzian
\begin{align}
\label{broadened:delta}
\Delta_\gamma(\omega) = \frac{1}{\pi} \frac{\gamma}{\omega^2+\gamma^2}
\end{align}
where we set $\gamma T^*=0.001$.

Figure~\ref{Box_Ix_noz} shows the nuclear spin noise calculated neglecting any nuclear Zeeman splitting ($z\equiv 0$) for different magnitudes of the transversal magnetic field. 
Since the spectra are even functions of $\omega$, we plot only the region $\omega\geqslant 0$. 
%
At $h\to 0$ a single Lorentzian peak is observed at $\omega=0$ in agreement with Eq.~\eqref{auto:h=0}: the electron impact on nuclear spins is negligible. With increasing $h$, 
the peak shifts away from zero frequency. 
The total spin $\mathbf F$ is no longer conserved in the presence of magnetic field,
and the Knight field generated by the precessing electron spin
results in the nuclear spin precession. At high magnetic field $h\gg 1$ the peak in the spectrum approaches $\omega^* = a_0/2$ [$\omega^* T^\ast=1/(2\sqrt{N})$],
originating
in the Knight field created by the electron spin.

\begin{figure}[t]
\begin{center}
\includegraphics[width=0.49\textwidth,clip]{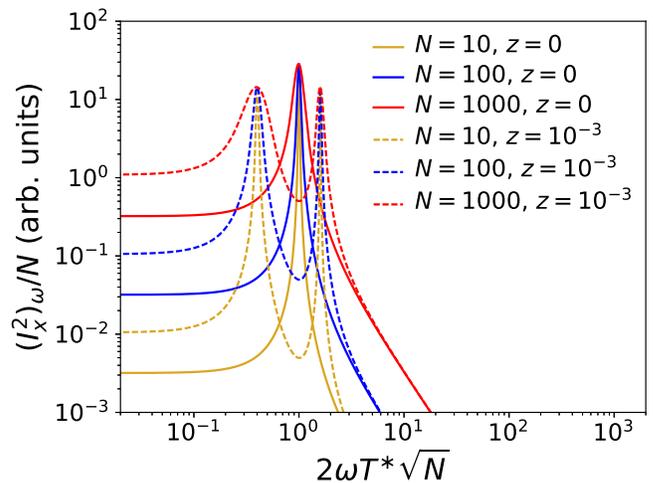}
\caption{The nuclear spin noise for different bath sizes, with and without nuclear Zeeman splitting, obtained by the Fourier transformation of Eq.\ \eqref{IxIx_box}. The external magnetic field is varied with the bath size, $h\sqrt{N}=300$ being kept constant. The spin noise is scaled with $N$ in order to make the plots comparable.}
\label{Nvgl}
\end{center}
\end{figure}

The effect of the nuclear Zeeman splitting is illustrated in Fig.~\ref{Box_Ix_z},  which shows the nuclear spin noise 
obtained from the Fourier transformation of Eq.\ \eqref{IxIx_box}. We assume that the nuclear Zeeman splitting is three orders of magnitude smaller (${z}=0.001$) than the electron Zeeman splitting. This ratio is typical for the experimentally studied systems~\cite{2015arXiv150605370B}. For relatively weak magnetic fields,
$zh \ll a_0$, 
the nuclear spin noise spectra are independent
of the nuclear Zeeman splitting. The influence of the nuclear Zeeman term 
becomes apparent for $zh>1/\sqrt{N}$ as a shift of the peak. At $h=10$, a double peak structure around $\omega=a_0/2$ emerges 
with the distance between the peaks governed by nuclear Zeeman splitting. 
Upon further increasing of the field, $zh \gg a_0$, 
the spin noise spectrum exhibits two closely lying peaks centered at $\omega = zh$, where the peak splitting is given by the hyperfine coupling constant $a_0$.

The nuclear spin noise dependency on the different bath sizes 
is illuminated in Fig.~\ref{Nvgl}. 
Here, the Fourier transformation of Eq.\ \eqref{IxIx_box} is shown for 
a constant combination of the external magnetic field and the bath size, $h\sqrt{N}=300$.  
Since the integral of the bath noise spectra results in $N/4$, the spin noise is scaled with $N$ in order to make the plots comparable. The peak location given by
the scale of $1/2T^{\ast}\sqrt{N}$ is not affected by the bath size. As the energy spectrum broadens with an increasing number of spins, so does the bath noise spectrum. Note that the characteristic time scale, $T^{\ast}$, depends on $N$.

The expression in Eq.~\eqref{IxIx_box} can be evaluated in $\mathcal{O}(N^2)$ computation time. This is a massive advantage to a full exact diagonalization, which is computed in $\mathcal{O}(2^{2N+2})$ or the Lanczos method, 
which computes the autocorrelation function in $\mathcal{O}(2^{N+1})$. 
Replacing the exponential
runtime by a quadratic one
allows for simulation of much larger systems. 
The disadvantage of this method is that it cannot 
include a distribution of coupling constants,
and it is impossible to extend the system to additional interactions if necessary. 
Below we briefly discuss the role of the distribution of the hyperfine coupling constants within the high magnetic field limit and 
in Sec.~\ref{sec:full} we will present the results within the numerical approach for arbitrary $h$ with account for the distribution of the coupling constants $a_k$.

\subsection{High magnetic field limit}

We have shown in Eq.~\eqref{auto:h=0} that the total nuclear spin correlation function does not decay in the box model in the large $N$ limit
and the absence of an external magnetic field. Decoherence is caused by a electric quadruple nuclear
interaction with random easy axis \cite{Bulutay2012,PhysRevLett.109.166605,PhysRevLett.115.207401}
as well as dipole-dipole interactions \cite{PhysRevB.74.195301, PhysRevB.74.035322, PhysRevLett.102.057601}
between neighboring nuclei. 
Here we
focus on the influence of the distribution 
of hyperfine coupling constants, which is clearly the dominating factor.
Before presenting general results, it is instructive to analyze 
first the limit of large magnetic fields, where $h\gg a_0$, but
with the requirement of the validity of the high-temperature limit:
the thermal orientation of electron and nuclear spins can be disregarded. 
At large external magnetic fields, electron-nuclear spin flips are suppressed, leading to the 
effective Hamiltonian
\begin{align}
H_{\text{hmf}} = hS_z + h\sum_k  z_k I_z + \sum_k a_k I_zS_z.
\label{H_hmf}
\end{align}
This Ising-like coupling is also applicable to some extent 
in the case of quantum dots with heavy holes~\cite{Testelin2009,Glazov2012,Chekhovich:2013ys} in an arbitrary magnetic field.
Without the spin-flip term, the Hamiltonian~\eqref{H_hmf} is already diagonal. In this case,
$\langle I_z(t) I_z(0)\rangle$ is given by Eq.~\eqref{auto:h=0} and 
\begin{align}
\begin{split}
\braket{I_x(t)I_x(0)}=& \frac{1}{4Z}\sum^{N}_{\mathclap{\substack{k=1,\\ L_k=\uparrow, \downarrow , \sigma=\uparrow, \downarrow}}} \braket{\sigma, \{L_k\}|I_k^x(t)I_k^x{(0)}|\sigma, \{L_k\}}\\
=& \frac{N}{4} \cos(a_0t/2)\cos(zht). 
\label{auto:h=infty}
\end{split}
\end{align}
Thus the nuclear spin noise spectrum contains four peaks at $\pm(a_0/2\pm zh)=\pm a_0(1/2\pm zh\sqrt{N})$. In the absence of a nuclear Zeeman splitting, the nuclear spin bath precesses in the magnetic field induced by the electron, the Knight field of the strength $a_0$. This distinguishes the characteristic frequency of the nuclear spin noise from those of 
the electron spin noise, which precesses in the Overhauser field with the characteristic
frequency $1/T^{\ast}$.

Noteworthy is that neglecting the spin-flip terms 
allows to extend the theory to arbitrary distributions of the hyperfine coupling constants $a_k$ and nuclear magnetic moments, 
so that we generalize Eq.~\eqref{auto:h=infty} to
\begin{equation}
\label{auto:h=infty1}
\braket{I_x(t)I_x(0)}= \frac{1}{4}\sum_{k=1}^N \cos(a_kt/2)\cos(z_kh t).
\end{equation}

\begin{figure}[tbp]
\begin{center}
\includegraphics[width=0.49\textwidth,clip]{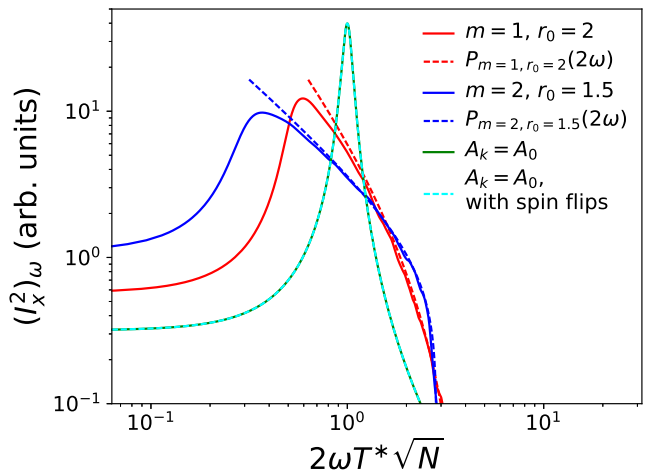}
\caption{The nuclear spin noise obtained from the Fourier transformation of Eq.\ \eqref{auto:h=infty1} at the high magnetic field limit for Ising coupling (continuous lines) and a bath size of $N=1000$. The nuclear Zeeman splitting is set to zero. Plotted are three different hyperfine coupling distributions, one with constant couplings and another one following Eq.\ \eqref{eq:Akspread} with $m=1, r_0=2$ and $m=2, r_0=1.5$. For comparison, the spin noise with spin flips, $h=1000$, and constant $a_k\equiv a_0$ is added with a dashed light blue line. The $a_k$ distribution function $\mathcal P(2\omega)$ obtained via Eq.\ \eqref{PA} is also provided for $m=2, r_0=1.5$ (blue dashed line) and $m=1, r_0=2$ (red dashed line). The spin noise was averaged over 10 sets of $a_k$ distributions.}
\label{High_Ix_noz}
\end{center}
\end{figure}

\begin{figure}[tbp]
\begin{center}
\includegraphics[width=0.49\textwidth,clip]{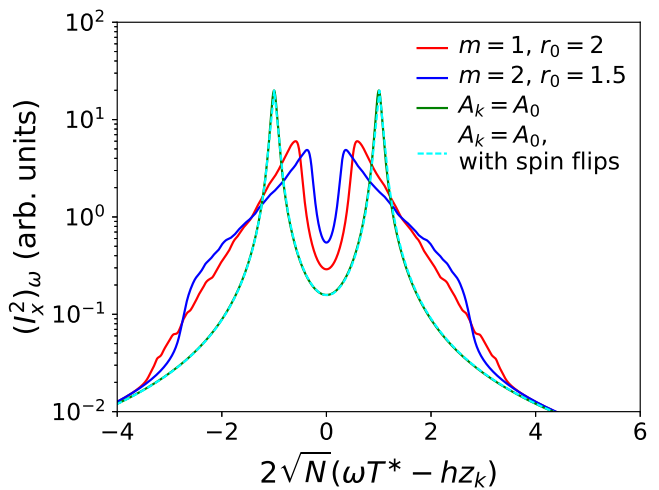}
\caption{The nuclear spin noise obtained from the Fourier transformation of Eq.\ \eqref{auto:h=infty1} at the high magnetic field limit with suppressed spin flips (continuous lines), $h=1000$, and a bath size of $N=1000$. The nuclear Zeeman splitting is $z_k=10^{-3}\,\forall\, k$. Plotted are three different hyperfine coupling distributions, one with constant couplings and another one following Eq.\ \eqref{eq:Akspread} with $m=1, r_0=2$ and $m=2, r_0=1.5$. For comparison, the spin noise with spin flips and constant $a_k\equiv a_0$ is added as a dashed line. The spin noise was averaged over 10 sets of $a_k$ distributions.}
\label{High_Ix_z}
\end{center}
\end{figure}

While the short-time dynamics of the spin system is not influenced by the particular form of the distribution of the hyperfine coupling constants, the distribution of $A_k$ plays an important role in the spin decoherence on the longer time scale, $t\gtrsim 10T^{\ast}$. 
In the high magnetic field limit
the summation over $k$ in Eq.\ \eqref{auto:h=infty1} can be performed analytically in the
large $N$ limit, and  nuclear spin noise spectrum acquires the form:
\begin{equation}
\label{NSN:high:spread}
(I_x^2)_\omega = \frac{{\pi}N}{4} \int da \mathcal P(a) \left[\Delta_\gamma(\omega-a/2)+\Delta_\gamma(\omega+a/2)\right],
\end{equation}
where $\mathcal P(a)$ is the distribution function of the coupling constants $a_k$. For 
a
quantum dot with an
envelope in $d$-dimensional space ($d=2$ for self-assembled quantum dots where the size quantization along the growth axis is much stronger than that in the quantum dot plane, $d=3$ for isotropic quantum dots) given by Eq.~\eqref{wavefunction},
$\mathcal P(a)$ can be written as by~\cite{PhysRevB.89.045317, Hack} 
\begin{equation}
\label{PA}
\mathcal P(a) =\frac{d}{m} \frac{1}{{r_0}^d a}\left[\ln{\left(\frac{a_{\rm max}}{a}\right)}\right]^{d/m-1}.
\end{equation}
Here, $r_0=R_0/L$ is the dimensionless cut-off parameter which characterizes the effective radius $R_0$ where the hyperfine coupling becomes unimportant, $a_{\rm max}$ is the maximal hyperfine coupling constant. 
We recall that $m$ is the power in the exponent of the  wave function stated in Eq.\ \eqref{wavefunction}:
$m=1$ for hydrogenic envelope function relevant for donor-bound electrons, $m=2$ for the Gaussian envelope functions. In our simulations we use the following set of parameters
\begin{align}
&m=1: \quad ~~{r_0}=2\\
&m=2: \quad ~~r_0=1.5,
\end{align}
in agreement with previous works \cite{PhysRevB.89.045317, Hack, PhysRevLett.115.207401}.

Figure~\ref{High_Ix_noz} shows the nuclear spin noise without the 
nuclear Zeeman-splitting, derived from the Fourier transformation 
of Eq.\ \eqref{auto:h=infty1}. In actual calculations it is convenient to generate first
a configuration of $A_k$ via
\begin{align}
A_k(x) = A_{\text{max}}\text{e}^{-r_0^m x^{m/3}},
\label{eq:Akspread}
\end{align}
where $x$ is the random real in the range $[0,1]$, 
and secondly normalize the couplings to $a_k=T^{\ast}A_k$. 
The time scale $T^{\ast}$, and therefore $a_{\text{max}}$, can vary slightly between different 
sets of the hyperfine couplings.
To stress that the box model including all spin flips 
perfectly matches the high magnetic field limit 
of the CSM with 
constant hyperfine couplings, both results are plotted. 
It follows from Eqs.~\eqref{broadened:delta} and \eqref{NSN:high:spread} that the nuclear spin noise spectrum
approaches
\begin{equation}
\label{NSN:high:spread:a}
(I_x^2)_\omega = \frac{{\pi}N}{2} \mathcal P(2\omega), \quad \omega>0
\end{equation} 
for $\gamma\to 0$.
To justify the use of the Ising model in the high magnetic field limit, the nuclear
spin noise calculated within the box model with (light blue dashed curve) and without (green curve) spin flips are included in Fig.\ \ref{High_Ix_noz} demonstrating perfect agreement. 
Analogous to the approach in the box model, the $\delta$-peaks 
of the nuclear spin noise was
broadened to a  Lorentzian with $\gamma T^*=0.001$.

Figure \ref{High_Ix_noz} also shows a comparison of the nuclear spin noise between the continuum limit of Eq.\ \eqref{NSN:high:spread:a} and the Fourier transformation of Eq. \eqref{auto:h=infty1} using 
the finite broadening 
for $N$=1000 nuclear spins in the absence of the nuclear Zeeman-splitting. 
The corresponding analytical curves calculated via Eq.~\eqref{NSN:high:spread:a} 
are plotted
in Fig.~\ref{High_Ix_noz} as dashed lines of the same color and fully agree with numerical calculations for $\omega>0.01$. 
The singularity  at $a=0$ of the distribution $\mathcal P(a)$, Eq.\ \eqref{PA},
is cut off by the dimensionless radius $r_0$ defining the smallest $a$ \cite{PhysRevB.89.045317}.
Therefore, the analytical and numerical solutions agree perfectly for the same  $\mathcal P(a)$ at high frequencies.
The difference for $\w\to 0$ arises from the hard cutoff of $\mathcal P(a)$ versus finite Lorenzian broadening
of the numerical solution
for a large but finite number of nuclei.

In Fig.~\ref{High_Ix_z} the nuclear spin noise is depicted  in the high magnetic field limit including the nuclear Zeeman term, again derived through the Fourier transformation of Eq.\ \eqref{auto:h=infty1}. 
The spin noise is strongly concentrated around $zh$, ($z_k=z\,\forall\,k$), for large external magnetic fields similar to the plot for $h=1000$ in Fig.\ \ref{Box_Ix_z}.
Again, the box model results with and without spin flips agree perfectly, 
which further underlines the validity of the use of the Ising model for 
high magnetic fields. 
The two peak structure centered at the nuclear Zeeman frequency $hz$ is clearly visible, similarly to the results plotted in Fig.~\ref{Box_Ix_z}. 
The curves  for the two distributions of $a_k$ are shown to illustrate
the additional broadening 
effect of the hyperfine coupling 
distribution onto the noise spectra and, consequently, the modified shape of peaks.

These results indicate the possibility of measuring the distribution of the hyperfine coupling constants directly via spin noise spectroscopy, as seen in Eq.\ \eqref{NSN:high:spread:a}.

\section{Full central spin model}\label{sec:full}

The full CSM cannot be easily solved analytically  at arbitrary magnetic fields
and with inhomogeneous hyperfine couplings although it is formally integrable via the Bethe ansatz~\cite{Gaudin1976}.
A quantum-mechanical, numerical approach based on the Lanczos method with restart (see the Appendix for details), can address this problem and 
also allows to incorporate additional interactions, such as the 
nuclear quadrupolar electric coupling \cite{PhysRevLett.109.166605,PhysRevLett.115.207401,PhysRevB.96.045441}.
In such an approach, however, the bath size is severely limited, since the Hilbert-space dimension increases exponentially with the number of nuclear spins. In this paper, $N=17$ spins are simulated, and the
$a_k$ distribution introduced in Eq.\ \eqref{eq:Akspread}
is used with a cut-off radius of $r_0=1.5$ and $m=2$. 
We perform a Fourier transformation
of the real-time dynamics to obtain the corresponding nuclear spin noise spectra;
a detailed description of the method can be found in the Appendix. 
Here, the nuclear spin correlation function is first evaluated in real time, and 
converted into a spectral function via a fast Fourier transformation. 
This differs from the previous approaches where the Fourier transformation was performed analytically.
In order to mimic larger bath sizes with the limited number of nuclear spins, we generated 25 different 
configurations of $N$ hyperfine couplings, $\{ a_k \}$, and perform a configuration average of the individual spectra.
We maintain the definition of the characteristic time scale $T^*$, Eq.\ \eqref{Tstar}, and state all
parameter in the previously introduced dimensionless units.

\begin{figure}[tbp]
\begin{center}
\includegraphics[height=0.34\textwidth,clip]{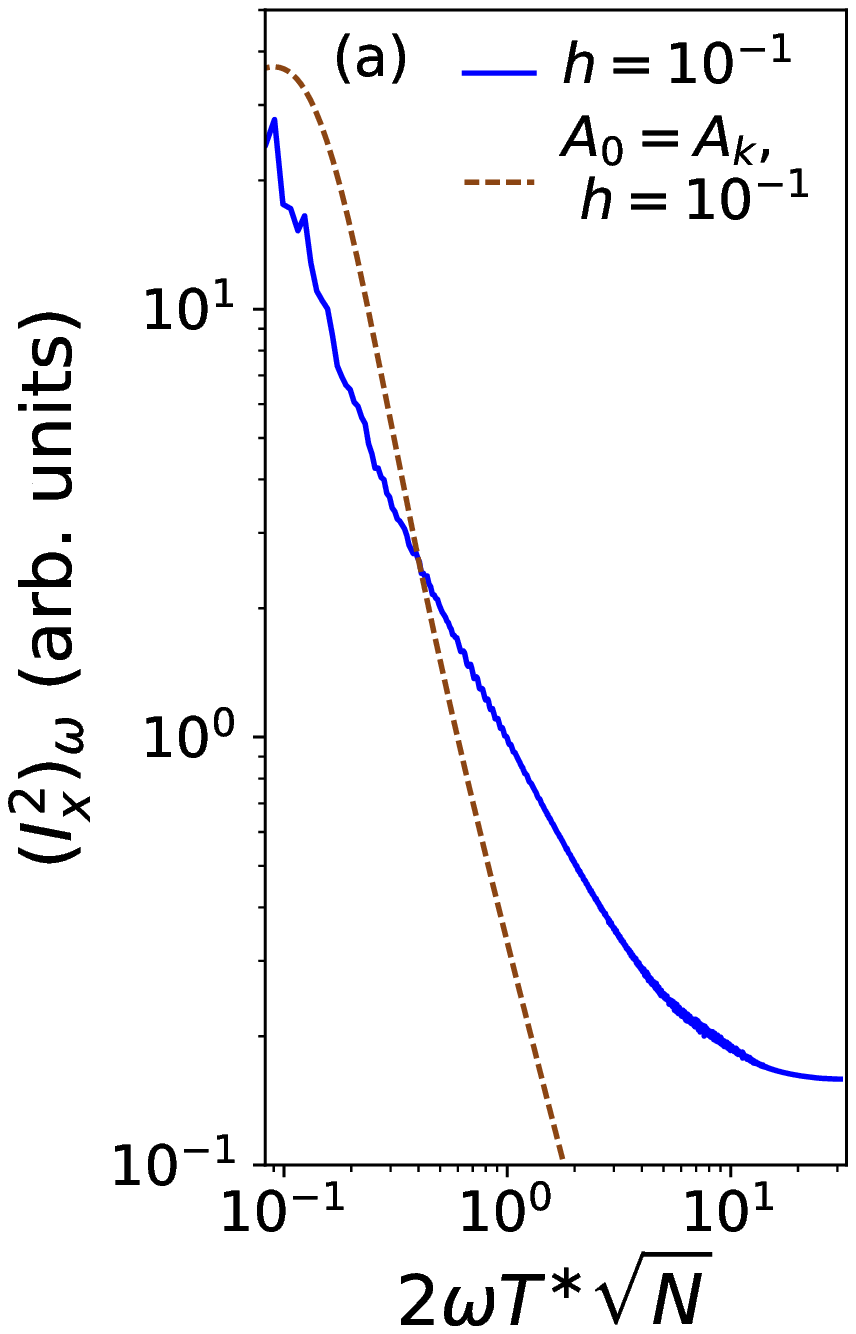}
\hspace{-10pt}
\includegraphics[height=0.34\textwidth,clip]{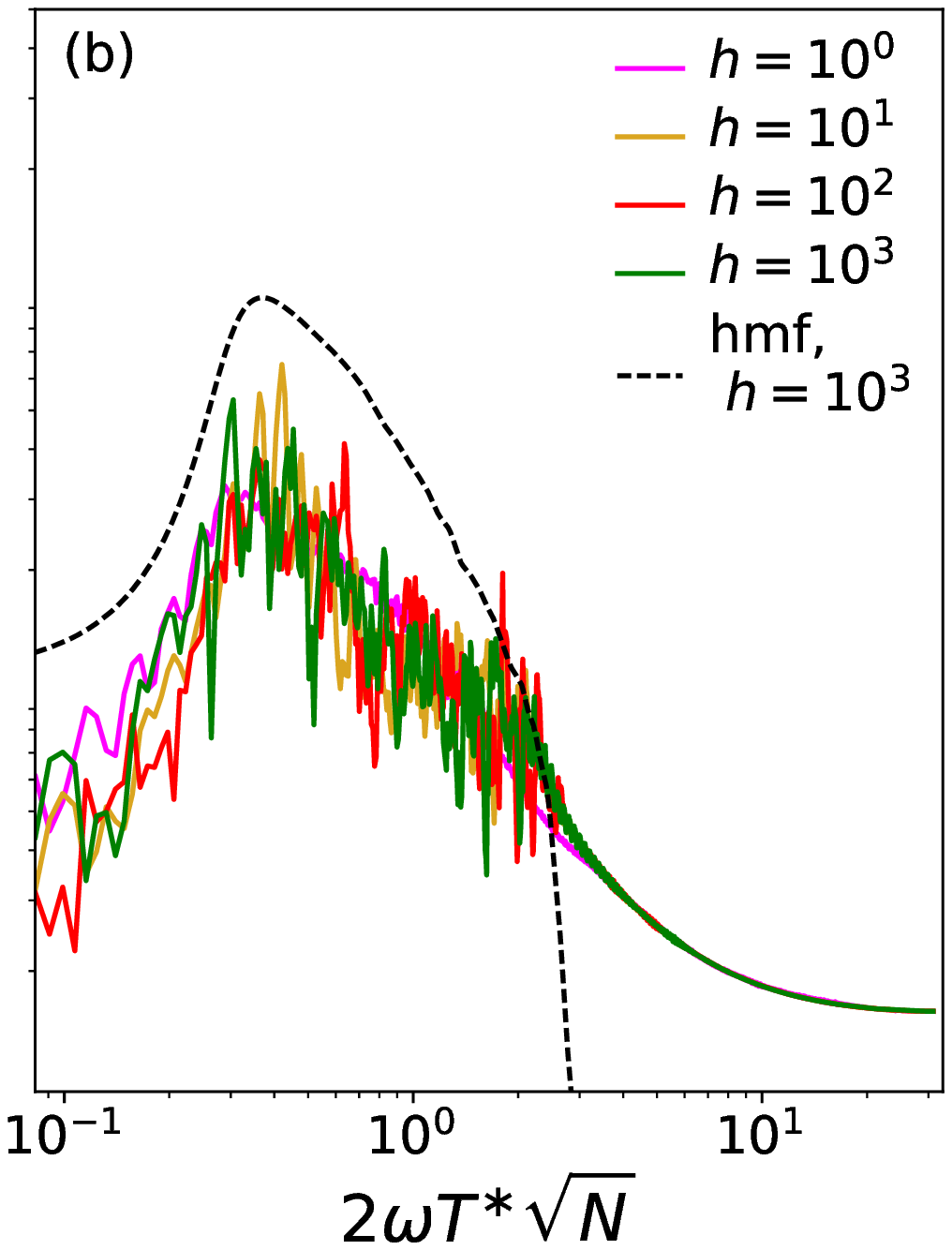}
\caption{The transversal nuclear spin noise computed with the Lanczos method 
for $N=17$ nuclear spins and a Krylov space dimension of $M=800$
in the absence of the nuclear Zeeman splitting. 
The $a_k$ distribution is based on a Gaussian electron wave function $m=2$. 
For better comparison the results for a low magnetic field are shown in (a) as well as the spectra obtained for the box model at a small magnetic field of $h=0.1$ (in a brown dashed line). In (b) the spin noise for higher magnetic fields is depicted, along with the
high magnetic field limit in a black dashed line.}
\label{fig:lan_noz}
\end{center}
\end{figure}

\begin{figure}[tbp]
\begin{center}
\includegraphics[height=0.35\textwidth,clip]{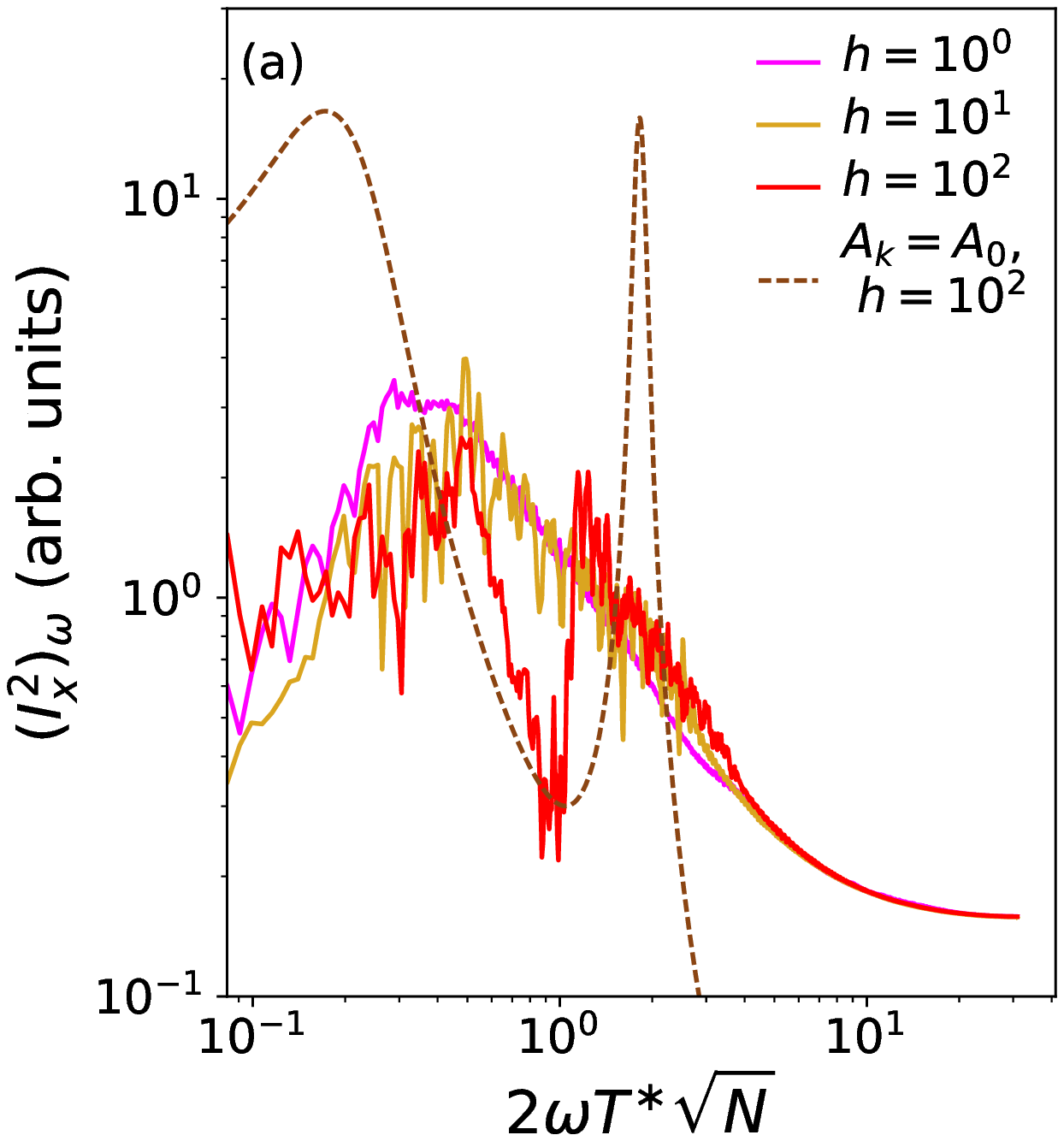}
\hspace{-10pt}
\includegraphics[height=0.35\textwidth,clip]{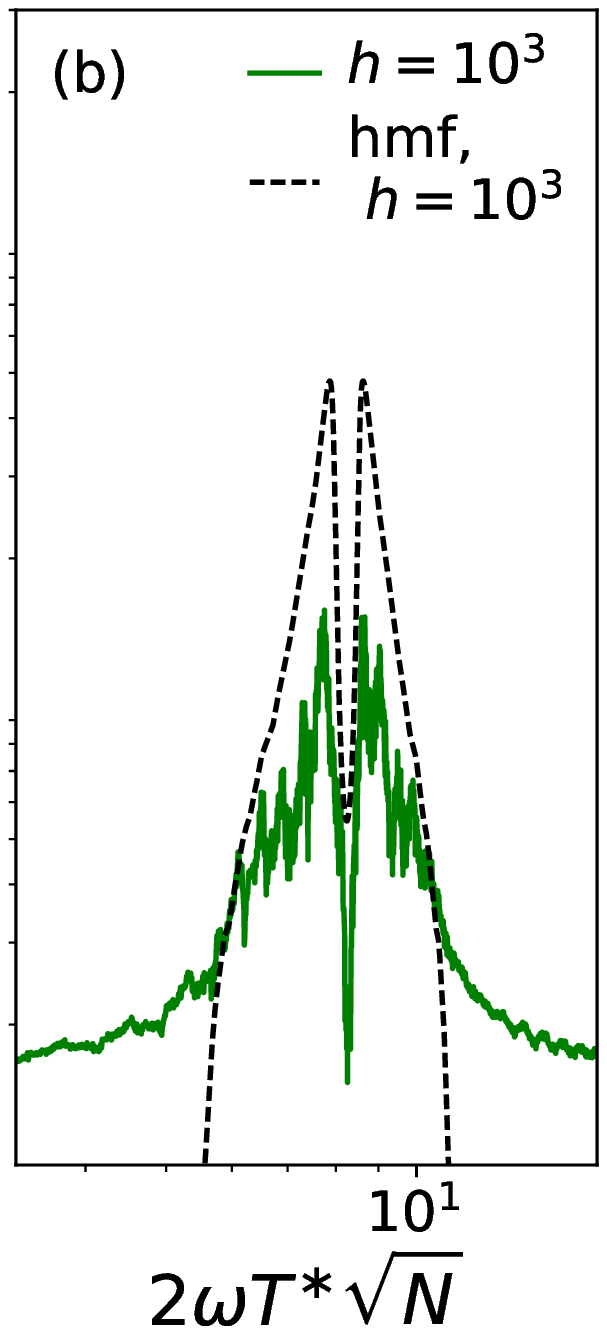}

\caption{The transversal
nuclear spin noise computed for the same parameters as in Fig.\ \ref{fig:lan_noz}
but including the electron Zeeman splittings by setting $z_k=z=10^{-3}\, \forall \, k$.
As with Fig.\ \ref{fig:lan_noz}, the low magnetic field spin noise is plotted in (a), while the high magnetic field spin noise, exemplary for $h=10^3$, is shown in (b).  Also depicted in (a) is the box model solution for $h=10^2$ in a dashed brown line. The high magnetic field limit is added to (b) for $h=10^3$ for comparison with the full model.
}
\label{fig:lan_z}

\end{center}
\end{figure}

Figures \ref{fig:lan_noz} (\ref{fig:lan_z}) show the 
nuclear spin noise spectra obtained by the Lanczos method without (with) nuclear Zeeman splitting in a transversal magnetic field. 

In Fig.~\ref{fig:lan_noz}(a) the spectrum for a small magnetic field ($h=0.1$) is depicted, along with the box model spectrum generated from Eq.\ \eqref{IxIx_box} for comparison.
At high magnetic fields (Fig.~\ref{fig:lan_noz}(b))
the spectra qualitatively resemble the previous analytical results obtained in high magnetic field limit
for $h>1$ and $\omega T^{\ast}<0.3$.
To illustrate that point, we added the analytical spectra 
computed analogously to the spectra in Fig.\ \ref{High_Ix_noz}
as dashed black curve using the same
$a_k$ distribution.
As shown in Eq.\ \eqref{auto:h=infty1}, the bath spin noise spectra center around the frequency of the Knight field in the absence of a nuclear Zeeman splitting. 
 Since the relative strength of the Knight field compared to 
 Overhauser field shrinks with increasing number of nuclear spins at a fixed $T^*$, 
the main spectral weight of the spin noise spectra
is located at a smaller frequency for $N=1000$ than the spin noise spectra for $N=17$.
We take this into account by rescaling the frequency axis with the factor $2T^{\ast}\sqrt{N}$.

The full numerical solution shows a much broader frequency spectrum than the analytical 
results for the box model in high magnetic fields.
But for $\omega T^{\ast}<0.2$ and $h>1$, the qualitative form of the 
nuclear spin noise in the full model is very well matched  by the high magnetic field limit.

Note that the finite number of nuclear spins in the Lanczos approach
 limits the number of discrete eigenvalues of the systems. 
Furthermore the fast Fourier transform does not include an additional 
Lorentz broadening. Both effects
lead to noisy spectra in comparison to the smooth spectra of the high magnetic field limit and the box model from Sec.\ \ref{sec:sns}.

A significantly different picture emerges for $h>10$
when the nuclear Zeeman term is included
in the Hamiltonian. 
The resulting transversal nuclear spin noise is shown in Fig.~\ref{fig:lan_z}(a) and (b)
for the same parameters as in Fig.~\ref{fig:lan_noz} but with a finite $z=10^{-3}$.
At $h=100$, in Fig.~\ref{fig:lan_z}(a), a clear drop becomes apparent at $hz=0.1$
whose origin becomes clear by the comparing the spectra with the
box model result added as a dashed brown line. 
The nuclear Zeeman splitting introduces a
two peak structure analytically illustrated by Eq.\ \eqref{auto:h=infty}. For a
low and an intermediate field, the 
shape of the spin noise spectrum is strongly influenced by the distribution function $ \mathcal P(a)$, and therefore, significantly
broader than the corresponding box model spectra.
The spectrum, however, approaches the analytical high magnetic field limit 
for $h=1000$ and  the same $a_k$-distribution -- 
plotted as a dashed black line for comparison in in Fig.~\ref{fig:lan_z}(b).

Comparing the results for the full CSM with the two analytical approximations,
the box model or the high magnetic field limit,
we can conclude that both analytical techniques give a good understanding of the basic features of the 
nuclear spin noise.
The analytical results can be obtained at very little computational costs and 
allow for bath sizes that are several orders of magnitudes larger than the bath sizes that can be treated in the quantum mechanical
evaluation of the general CSM.

\section{Conclusion}\label{sec:concl}

We studied the nuclear spin fluctuations,
both analytically and numerically, using the central spin model. 
This model includes the hyperfine interaction of nuclear spins with electron spin 
as well as the Zeeman effects for the electrons and the nuclei. 
For homogeneous coupling constants, i.e., within the box model approximation, the system is exactly solvable, and the auto-correlation function of the total bath polarization can be obtained
analytically for an arbitrary number of nuclear spins. 
Including the nuclear Zeeman splitting  in the model, we observe the emergence of two peaks in the 
nuclear spin noise spectrum centered at the nuclear spin precession frequency.
The distance between the peaks is given by the hyperfine coupling constant. 
The effect of the hyperfine coupling distribution function
constants was studied analytically
in the high magnetic field limit. 
We demonstrate that the shape of the nuclear spin noise spectrum is controlled by 
this distribution function
of the hyperfine coupling constants, i.e., by the shape of the electron envelope function.
A Lanczos algorithm with restarts was employed
to numerically obtain the bath correlation 
for inhomogeneous couplings and arbitrary magnetic fields. 
The agreement between the numerical and analytical solutions is particularly good at high magnetic fields. 
Therefore, the analytical results 
provide a good description 
of the system, but with significantly reduced computational costs.

\begin{acknowledgments}
We thank D.S. Smirnov for valuable discussions.
We acknowledge the financial support by the
Deutsche Forschungsgemeinschaft and the Russian Foundation of Basic
Research through the transregio TRR 160. M.M.G. is grateful to RFBR projects 17-02-00383 and 15-52-12012 and Russian President grant MD-1555.2017.2 for partial support.
\end{acknowledgments}

\appendix

\section{Lanczos Method with Restarts}
\label{appA}

We used the Lanczos time evolution method with restarts \cite{PhysRevB.96.045441} in order to keep stability and stochastic evaluation of the trace
\cite{RevModPhys.78.275,PhysRevB.89.045317}
by averaging over a small number $R \ll D$ of randomly chosen states $\ket{r}$. The relative error made by the stochastic evaluation of the trace
\begin{align}
\text{Tr}(O) \approx \frac{1}{R} \sum_{r=0}^{R-1} \braket{r|O|r}
\end{align}
is of the order $\mathcal{O}(1/\sqrt{DR})$ \cite{RevModPhys.78.275}. 
Using this technique, we approximate the second order nuclear spin auto-correlation function by
\begin{align}
\braket{I_x(t)I_x(0)} = \frac{2}{RD}\sum_r^{R}\braket{1_r(t)|I_x|2_r(t)}
\label{eq:IxIx_lan}
\end{align}
where
\begin{eqnarray}
\ket{1_r(t)} &=& e^{-\mathrm{i}H t} \ket{r} \\
\ket{2_r(t)} &=& e^{-\mathrm{i}H t} I_x \ket{r}.
\end{eqnarray}
The nuclear spin dynamics happen on a time scale much longer than the electron spin dynamics, we need to be able to simulate up to large times without losing numerical accuracy.
Therefore we discretize the time $t_n=n\tau, n=0,..., N_t$ to to calculate the evolution of the state as
\begin{equation}
\ket{\psi(t_n)} = e^{-\mathrm{i}H \tau} \ket{\psi(t_{n-1})} 
\end{equation}
in smaller time steps $\tau$.

The Lanczos-Krylov Algorithm \cite{Lanczos50aniterative} is used here for calculating
the real-time propagation. The vectors
\begin{align}
\ket{\psi}, H\ket{\psi}, H^2\ket{\psi}, \ldots, H^M\ket{\psi},
\end{align}
define the Krylov space, a $M$ dimensional subspace of the Hilbert space.
Each time, the state $\ket{\psi(t_{n-1})} $ from the previous time step is used
as new starting vector. Via the Lanczos algorithm, we gain the
$M\times M$ dimensional tridiagonal matrix representation of the Hamiltonian $\underline{\underline{H}}(M)$ in the Krylov space. We diagonalize $\doubleunderline{H}(M)$ and obtain the approximate eigenstates $\ket{\nu^n}$
\begin{align}
\underline{\underline{H}}(M)\ket{\nu^n}= \epsilon_\nu^n\ket{\nu^n}
\end{align}
and the corresponding eigenvalues $\epsilon_\nu^n$. The time evolution can then be approximately written as
\begin{align}
\ket{1_r(t_n)} = \mathrm{e}^{-\mathrm{i}H\tau}\ket{1_r(t_{n-1})}=\sum_{\nu}^M \mathrm{e}^{-\mathrm{i}\epsilon_\nu^n\tau}
c^1_n(t_{n-1})  \ket{\nu^n}.
\label{eq:1rtn}
\end{align} 
where $c^1_n(t_{n-1})  = \braket{\nu^n|1_r(t_{n-1})}$. The result becomes exact for $\tau\to 0$.
To arrive at $\ket{1_r(t_{N_t})}$, $N_t$ Lanczos time evolution steps have to be computed.

The same time evolution is needed for the vector $\ket{2_r(t)}$,
\begin{align}
\ket{2_r(t_n)} = \mathrm{e}^{-\mathrm{i}H\tau}\ket{2_r(t_{n-1})}=\sum_{\nu}^M \mathrm{e}^{-\mathrm{i}\epsilon_\nu^n\tau}
c^2_n(t_{n-1})
\ket{\nu^n},
\label{eq:2rtn}
\end{align} 
where $c^2_n(t_{n-1})=\braket{\nu^n|2_r(t_{n-1})}$. Having gained both $\ket{1_r(t_n)}$ and $\ket{2_r(t_n)}$, we can calculate the nuclear spin auto-correlation function $\braket{I_x(t)I_x(0)}$ using Eq.\ \eqref{eq:IxIx_lan}. The nuclear spin noise is then computed via fast Fourier-transformation. For $n$ time steps $2N_t$ Lanczos iterations are needed. A finer time resolution can be achieved by substituting $\tau$ with $\Delta t$ ($0<\Delta t<\tau$) in Eq.\ \eqref{eq:1rtn} and \eqref{eq:2rtn}, calculating the nuclear spin correlation function at $t_{n-1}+\Delta t$. This way, several time steps between $t_{n-1}$ and $t_n$ can be calculated without having to do the Lanzcos algorithm with a different starting vector, thus saving computation time. This gives us the flexibility to chose $\tau$ smaller or larger depending on the dynamics of the system, while keeping the same time resolution.

\end{document}